\documentclass[
 twocolumn,
 amsmath,
 amssymb,
 aps,
 prc,
 10pt
]{revtex4-2}

\usepackage{graphicx}
\usepackage{physics}
\usepackage[hidelinks]{hyperref}

\newcommand{\h}[1]{\hat{#1}}

\graphicspath{{figures/}}

\begin{document}

\title{Weak entanglement approximation for nuclear structure}
\author{Oliver C. Gorton} 
\affiliation{Lawrence Livermore National Laboratory, Livermore, California, USA}
\affiliation{San Diego State University, San Diego, California, USA}
\author{Calvin W. Johnson} 
\affiliation{San Diego State University, San Diego, California, USA}

\begin{abstract}
The interacting shell model, a configuration-interaction method, is a venerable
approach for low-lying nuclear structure calculations; but it is hampered by the exponential
growth of its basis dimension as one increases the single-particle space and/or
the number of active particles. Recent, quantum-information-inspired work has
demonstrated that the proton and neutron sectors of a nuclear wave function are
weakly entangled. Furthermore, the entanglement is smaller for nuclides away from
$N=Z$, such as heavy, neutron-rich nuclides. Here we implement a weak
entanglement approximation to bipartite configuration-interaction wave
functions, approximating low-lying levels by coupling a relatively small
number of many-proton and many-neutron states. This truncation scheme, which we
present in the context of past approaches, reduces the basis dimension by many
orders of magnitude while preserving essential features of nuclear spectra.
\end{abstract}

\maketitle

\section{Introduction}\label{sec:intro}

In 1949, Haxel, Jensen, and
Suess~\cite{PhysRev.75.1766.2} and Goeppert-Meyer~\cite{PhysRev.75.1969}
introduced the non-interacting shell model for nuclear levels. A decade later,
Kurath~\cite{PhysRev.101.216}, followed by Halbert and
French~\cite{PhysRev.105.1563} among others, introduced configuration mixing,
also known as the configuration-interaction method or shell-model
diagonalization. In configuration-interaction, one expands the nuclear wave
function in a basis of many-body states, 
\begin{equation}
    | \Psi \rangle = \sum_\alpha c_\alpha | \alpha \rangle, \label{CI}
\end{equation}
computes the matrix elements of the nuclear Hamiltonian $\hat{H}$ in that basis,
$H_{\alpha \beta}= \langle \alpha | \hat{H} | \beta \rangle$, and then solves
the matrix eigenvalue problem,
\begin{equation}
    \sum_\beta H_{\alpha \beta}c_\beta = E c_\alpha.
\end{equation}
In the 1970s, Whitehead \textit{et al.} introduced the Lanczos algorithm to the
interacting shell model~\cite{whitehead1977computationala}, allowing one to
efficiently find extremal eigenstates in basis dimensions for which full
diagonalization would be prohibitive. On today's supercomputers one can tackle
basis dimensions in the few tens of
billions~\cite{forssen2018large,mccoy2024intruder}.

Even modern supercomputers are not enough, however. The many-body basis
dimension goes like $N_s$ choose $N_p$ $ = N_s!/ N_p! (N_s -N_p)!$, where $N_s$
is the number of available single-particle states and $N_p$ is the number of
active particles. (Selection rules reduce these dimensions but the scaling
remains the same.) As the factorial leads to an unfavorable exponential scaling,
this motivates alternate methods, such as coupled
clusters~\cite{PhysRevC.82.034330}, which scale polynomially rather than
exponentially. 

Nonetheless, configuration-interaction methods see continued development due to
their numerous advantages including: relative ease of generating excited states,
ability to handle even and odd numbers of particles equally well, relevance to
open-shell nuclides, flexibility with choice of interactions, and so on.
Examples of alternate truncations include the so-called Monte Carlo Shell
Model~\cite{otsuka2001monte}, truncations based upon algebraic
structures~\cite{launey2014emergence,PhysRevLett.125.102505},  energy-based
importance truncations~\cite{PhysRevC.93.021301}, and beyond-mean-field-based
truncations~\cite{PhysRevC.105.054314}.

A surge in the interest in and applications of quantum information theory has
given new lenses for truncation schemes. The nucleus can be naturally described
as a bipartite systems with proton and neutron components, and nearly all
shell-model codes utilize such a
partitioning~\cite{johnson2013factorization,shimizu2013nuclear,shimizu2019thick}.
Recent work has indicated that the proton and neutron components are only weakly
entangled~\cite{johnson2023protonneutron}, and in fact among different
partitioning schemes, proton-neutron partitioning leads to the smallest
entanglement~\cite{perez-obiol2023quantum}. 

We use these observations to motivate our \textit{weak entanglement
approximation}, where our lowest-order calculation, generating the proton and
neutron bases independently, implicitly assumes zero entanglement. This approach
can be related to density-matrix renormalization group
calculations~\cite{white1992density, white1993densitymatrix,
papenbrock2005density} as well as the singular-value-decomposition variational
wave function~\cite{papenbrock2004solution, papenbrock2003factorization}
approaches. Our Proton And Neutron Approximate Shell-model (PANASh) is, however,
more straightforward than either of those approaches, as we do not iterate to
optimize the basis. Nonetheless, as we demonstrate here, our simplified protocol
provides a very good description of nuclear spectra for a variety of different
cases. 

In Section~\ref{sec:entangle} we briefly review entanglement of bipartite
systems.  In Section~\ref{sec:panash} we outline the PANASh scheme.  We then
present results: in Section~\ref{sec: benchmark} we compare against full
configuration interaction cases, while in Section~\ref{sec:heroic} we
demonstrate the utility of PANASh by presenting an application too large to
tackle in the standard shell model. After comparing in Section~\ref{sec:compare}
to related methods which rely upon singular-value-decomposition, we briefly
outline work yet to be done. In Appendix~\ref{sec:appendix} we give details
of coupling together the proton and neutron components.

\section{Proton-neutron entanglement}\label{sec:entangle}

In this section we briefly review the theory of entanglement and what we mean by
`weak' entanglement. Entanglement starts by considering two independent Hilbert
spaces, which here we write as $\mathcal{P}, \mathcal{N}$ to reflect the proton
and neutron spaces relevant to the shell model. One then constructs a product
Hilbert space $\mathcal{H} = \mathcal{P} \otimes \mathcal{N}$. This can be done
explicitly by writing basis states $\{ | \alpha \rangle \}$ of $\mathcal{H}$ as
simple tensor products of basis states from the component spaces, 
$\{ \ket{p} \} \in \mathcal{P}$ and 
$\{ \ket{n} \}\in \mathcal{N}$, 
so that each $| \alpha \rangle = \ket{p} \otimes \ket{n}$. 
The dimension of the bipartite Hilbert space $\mathcal{H}$ is multiplicative:
\begin{equation}
    \dim{\mathcal{H}}  = \dim{\mathcal{P}} \times \dim{\mathcal{N}}.
\end{equation}
(In practice quantum number selection rules, such as on total $J_z$, can make
the Hamiltonian block-diagonal and thus reduce the working dimension.) For
configuration-interaction (CI) calculations, one then simply expands in the
basis as in Eq.~(\ref{CI}), or, explicitly representing the bipartite nature of
the space, 
\begin{equation}\label{eq:generalpnwfn}
    | \Psi\rangle = \sum_{p,n} \psi_{p,n}  \ket{p} \otimes \ket{n}.
\end{equation}

Any interaction which couples two systems will generate an entangled state. An
entangled state can no longer be written as a simple product
$\ket{\Psi}=\ket{\tilde{p}}\otimes\ket{\tilde{n}}$, but will necessarily involve
a superposition (sum) of product states, as in Eq.~\eqref{eq:generalpnwfn}. This
relationship can be formalized using density operators. A system with a density
operator $\hat\rho$ is said to be entangled if its von Neumann entropy,
\begin{equation}\label{eq:vne}
    S(\hat\rho) = -\tr(\hat\rho \ln \hat\rho),
\end{equation}
is nonzero. The density operator for any wave function is $\hat{\rho} = | \Psi
\rangle \langle \Psi |$ and the elements of the density matrix are 
\begin{equation}
    \rho_{p n,p^\prime n^\prime} = \psi^*_{pn} \psi_{p^\prime n^\prime}.
\end{equation}
We often compute Eq.~\eqref{eq:vne} using the eigenvalues of $\hat\rho$. For an
isolated system, the density matrix has unit trace if the wave function is
normalized, and as the density matrix is idempotent ($\hat{\rho}^2 =
\hat{\rho}$) the eigenvalues of the density matrix are either 0 or 1. Therefore,
its entanglement entropy $S(\rho)$ is identically 0. However, we can consider
the entanglement between partitions within our isolated system which are
interacting. This is done using \textit{reduced density matrix} which is found
by tracing over one of the subspaces: $\rho^\text{(x)} = \tr_y\rho^{(xy)}$. For
example, tracing over the neutron partition yields a reduced density matrix with
proton indices:
\begin{equation}
    \rho^\mathrm{(p)}
    _{p,p^\prime} = \sum_n \rho_{pn,p^\prime n} = \sum_n \psi^*_{pn} \psi_{p^\prime n}.
\end{equation}
The reduced density matrix still has unit trace, but its eigenvalues can be
between 0 and 1, inclusive. This happens whenever there is an interaction
coupling the two partitions. If any eigenvalues are not zero or one, the two
partitions are \textit{entangled}, and the entropy $S(\hat{\rho}^x) > 0.$

By the singular-value decomposition theorem, the eigenvalues of the reduced
density matrix do not depend upon which partition index is summed over. Indeed,
one can simply view entanglement through the lens of singular-value
decomposition (SVD), also called Schmidt decomposition. The proton-neutron
coefficients of equation~\eqref{eq:generalpnwfn} form a matrix on which we can
perform a SVD: $\Psi = USV^T$, which can be seen as a transformation of the
proton and neutron basis factors using the orthogonal matrices $U$ and $S$ to
one in which the $\Psi$ matrix is diagonal ($S$). This is equivalent to using
the eigenvectors of the proton density matrix $\rho^{(p)}=US^2U^T$ as the proton
basis factors and the eigenvectors of $\rho^{(n)} = VS^2V^T$ as the neutron
basis factors; simultaneously we see that the eigenvalues of the reduced density
matrix are nothing but squares of the singular values. 

Now we can summarize the relation between entanglement, SVD, and wave function
factorization: a state with zero entanglement has singular values of only 1's
and 0's, and thus can be written as a single term, a simple product of one
proton factor and one neutron factor. Not coincidentally, both density matrix
renormalization group calculations~\cite{white1992density,
white1993densitymatrix, papenbrock2005density} and variational wave function
truncations~\cite{papenbrock2004solution, papenbrock2003factorization} rely upon
SVD.

In recent work~\cite{johnson2023protonneutron} we found empirical evidence that
low-lying shell model states have weak proton-neutron entanglement, that is,
entanglement much smaller than the maximum; furthermore nuclei with $N>Z$ have
systematically lower entanglement that their $N=Z$ counterparts. These results
suggest that a truncation scheme based on proton-neutron factorization may be
even more effective for neutron-rich nuclei, where the need for reduced
dimensions is greatest. It has also been shown that among orbital
equipartitions, the proton-neutron bipartition has the weakest
entanglement~\cite{perez-obiol2023quantum}. In the limit of zero entanglement,
an eigenstate can be written as a simple tensor product  of one proton wave
function times one neutron wave function. In such a case, the effective
dimension of the joint model space would be greatly reduced to an additive one:
\begin{equation}
    \lim_{\text{entanglement} \to 0}\dim{\mathcal{H}}  
    = \dim{\mathcal{P}} + \dim{\mathcal{N}}.
\end{equation}
It is this effective reduction of dimensionality that drives our method. We
surmise that our method should therefore out-perform other orbital-partitioning
truncation schemes such as~\cite{andreozzi2001redundancyfree}.

In this paper we present the proton and neutron approximate shell model which
builds upon the weak entanglement limit. In simple terms, we first solve the
Hamiltonian in the zero-entanglement limit (setting $\h H^{(pn)}$ to zero). This
leaves us with the uncoupled proton and neutron wave functions which we call the
proton and neutron factors. Second, we couple these factors together, now in all
combinations suitable to form basis states with good total angular momentum. The
number of factors from each subspace is truncated to suit a chosen reduction in
overall basis dimension (generally limited by computer resources). Third, the
full interaction including $\h H^{(pn)}$ is diagonalized in this truncated
basis.

\section{Proton and neutron approximate shell model}\label{sec:panash}

We now explain how the weak entanglement limit is used with a proton-neutron
factorization to approximate exact shell model states. The shell model
Hamiltonian represents a system of interacting single-particle
harmonic oscillator states with a mean-field (one-body) and effective two-body
interaction:
\begin{align}\label{eq:hsingle}
    \h H = \sum_{i} \epsilon_i \h a^\dagger_i \h a_i
    + \frac{1}{4} \sum_{ijkl} V_{ijkl} \h a^\dagger_i \h a^\dagger_j \h a_k \h a_l.
\end{align}
The creation/destruction operators $\h a_i^\dagger$/$\h a_i$ create/destroy
particles in the valence space orbital $i$, which has harmonic oscillator labels
$n_i$, $l_i$, $j_i$ (and magnetic quantum number $m_i$ for single-particle
states).

With two species of particles, protons and neutrons, we have the following
Hamiltonian:
\begin{align}\label{nucham}
    \h H = \h H^{(p)} + \h H^{(pp)} + \h H^{(n)} + \h H^{(nn)} + \h H^{(pn)},
\end{align}
where the superscript in parenthesis indicates the type of operator: $(p)$ is a
one-body proton operator, $(pp)$ is a two-body proton operator, and equivalently
for neutrons $(n)$, $(nn)$; and finally there is the remaining proton-neutron
two-body interaction $(pn)$.  The proton-only and neutron-only operators,
\begin{align}
    \hat{{P}} &\equiv \h H^{(p)} + \h H^{(pp)}\\
    \hat{{N}} &\equiv \h H^{(n)} + \h H^{(nn)},
\end{align}
each have the form of Eq.~\eqref{eq:hsingle}. Furthermore, each is an operator
constrained to its own subspace: $\h P: \mathcal{P} \to \mathcal{P}$ and $\h N:
\mathcal{N} \to \mathcal{N}$. The direct-product of these two subspaces is the
bipartite proton-neutron space $\mathcal{H} = \mathcal{P} \otimes \mathcal{N}$,
which is where the total Hamiltonian acts:
\begin{equation}
    \h H = \h P + \h N + \h H^{(pn)}.
\end{equation}
Each subspace operator has its own eigenstates and eigenenergies:
\begin{align}\label{peig}
    \h P |p \rangle &= E_p |p\rangle \\
    \label{neig}
    \h N |n\rangle &= E_n |n\rangle.
\end{align}
The dimensions of these subspaces are orders of magnitude smaller than the full
space, and often can be solved without any truncation. In the weak entanglement
limit, these subspace eigenstates approximate the optimal basis factors
$\ket{\tilde p_j}, \ket{\tilde n_j}$, i.e. eigenstates of the exact reduced
density matrix. This motivates us to use these proton and neutron eigenstates
directly as factors for a basis:
\begin{align}\label{base}
    [|p\rangle \otimes  |n\rangle]_{J^\pi} = |pn; J^\pi\rangle.
\end{align}
We work in the J-scheme so that any truncation of this basis will produce wave
functions with well-defined $J$. (This is similar to the methodology of the
$J$-scheme (fixed total $J$) configuration-interaction code 
{\tt NuShellX}~\cite{brown2014shell}, except, crucially, we carry out an energy
truncation on the proton and neutron components.)

We can write the matrix elements of the Hamiltonian in this basis as
\begin{align}\label{eq:panashham}
    \bra{p_f n_f; J^\pi} \hat H \ket{ p_i n_i ; J^\pi} 
    = \delta_{n_f n_i} E_p + \delta_{p_fp_i} E_n \nonumber \\
     + \bra{p_f n_f; J^\pi} \hat H^{(pn)} \ket{ p_i n_i ; J^\pi},   
\end{align}
where the matrix elements of $\hat H^{(pn)}$ are expressed in terms of one-body
density matrices computed from the $\ket{p}$ and $\ket{n}$ eigenstates. The
details for computing the proton-neutron matrix elements are given in
Appendix~\ref{sec:appendix}. By diagonalizing in a truncated basis set, we
obtain approximate solutions of the form:
\begin{align}\label{approxsol}
     |\tilde{\Psi}\rangle = \sum_{pn}^{m_n,m_p} \psi_{pn} |pn; J^\pi\rangle 
     \approx |\Psi\rangle,
\end{align}
where $|\Psi\rangle$ is an exact eigenstate of $\hat H$. Setting  $m_p =
d_p\equiv \dim(\h P) $ and $m_n = d_n\equiv \dim(\h N)$ leads to the full
configuration interaction solution. In the next section we show we can obtain
good results with $m_p \ll d_p, m_n \ll d_n$; the  energies follow the usual
variational principle.

\section{Results\label{sec: benchmark}}

Here we compare the low-lying spectra obtained from the weak entanglement
factorization against  untruncated, full configuration-interaction (FCI)
calculations.  Given specifications from the phenomenological interactions used,
we also compute the total binding energies using the formulas given
in~\cite{cole1999predicted, honma2009new, honma2004new}. We compare the
most-bound levels for four benchmark nuclei, $^{78}$Ge, $^{70}$As, $^{60}$Ni,
$^{79}$Rb. In each case, multiple calculations are performed with increasing
fidelity: each uses an increasing fraction of the proton and neutron subspace
factors and therefore an increasing computational cost which scales like the
cube of the model space dimension.  

We perform calculations in two model spaces. The first is the $0f_{7/2},
1p_{3/2}, 0f_{5/2}, 1p_{1/2}$ space with the GX1A
interaction~\cite{honma2004new}, and the second is the $0f_{5/2}, 1p_{3/2},
1p_{1/2}, 0g_{9/2}$ space with the JUN45 interaction~\cite{honma2009new}. The
nuclei modeling in each space and their dimensions are shown in
Table~\ref{tab:panashcalcs}.
\begin{table*}[htb]
    \centering
    \caption[PANASh benchmark calculations]{PANASh benchmark calculations and
    their: interaction used (Int.), $M$-scheme FCI dimension in millions (Mdim),
    number of protons ($Z$), number of valence protons ($Z_\text{val.}$), proton
    subspace $M$-scheme dimension (Zdim), (the equivalent for neutrons), and
    properties for which the nucleus was selected as a benchmark.}
    \label{tab:panashcalcs}
    \begin{tabular}{c c c c c c c l}
    \hline
        Nucleus & Interaction & M-scheme   & $Z$ $(Z_\text{val.})$ & Z dim. & $N$ $(N_\text{val.})$ & N dim. & Properties \\
        & & FCI dim. ($\times10^6$) & & & & & \\
    \hline
        $^{78}$Ge & JUN45 & 3.7 & 32 (4)& 701    & 46 (18) & 701    & even-even, deformed \\
        $^{70}$As & JUN45 & 760 & 33 (5)& 2\,293  & 37 (9)  & 36\,998 & odd-odd, deformed \\
        $^{60}$Ni & GX1A  & 1\,090 &28 (8)& 12\,022 & 32 (12) & 12\,022 & even-even, spherical \\
        $^{79}$Rb & JUN45 & 8\,600 &37 (9)& 36\,998 & 42 (14) & 24\,426 & odd-A, spherical \\
    \end{tabular}
\end{table*}
These nuclei were selected to span a large range of $M$-scheme model space
dimensions (from $10^6$ to $10^9$), as well as several properties which affect
the difficulty of capturing the many-body physics. Even-even nuclei have more
regular, collective excitation spectra than odd-A ($^{79}$Rb) or odd-odd
($^{70}$As) nuclei.
We consider two even-even cases: one more spherical ($^{60}$Ni) expected to
exhibit seniority-like spectra, and one more deformed ($^{78}$Ge)  expected to
exhibit rotational spectra. 

Since PANASh uses a $J$-scheme basis, each $J^\pi$ block of the Hamiltonian can
be solved independently with a much smaller basis, typically an order of
magnitude than the equivalent $M$-scheme basis. The $J$-scheme matrix elements
have a much higher cost per element, however, and the {\tt BIGSTICK} code is
more efficient for a fixed-size basis. For this reason we use {\tt BIGSTICK} to
compute the FCI results for the un-truncated basis where the advantages of the
weak-entanglement approximation are lost.

\begin{figure*}[htb]
    \centering
    \includegraphics[width=.65\textwidth]{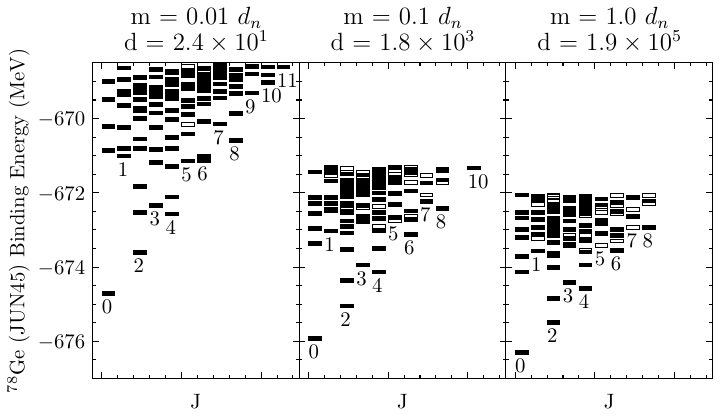}
    \caption[$^{78}$Ge binding energy]{Binding energies of 100 levels of
    $^{78}$Ge computed with the weak entanglement approximation. Each panel has
    two labels: the number of components used as a fraction of the neutron
    subspace dimension, and the largest $J$-scheme basis dimension required
    (across all values of $J$). Each stack of bars is a set of levels with a
    given total angular momentum $J$; full bars are positive parity and empty
    bars are negative parity.} 
    \label{fig: ge}
\end{figure*}

The results for $^{78}$Ge, an even-even deformed nucleus, are shown in
Fig.~\ref{fig: ge}.
In each panel the PANASh basis is constructed with $m$ proton and neutron
factors as an increasing fraction of the available subspace factors
$d_n=d_p=701$ (see also Table~\ref{tab:panashcalcs}).  Also given is the maximum
$J$-scheme dimension $d$ solved. In the first panel, with $1\%$ of the basis
factors resulting in four orders of magnitude reduction in the dimension of the
model space, we reproduce the spectral structure characteristic of a deformed,
rotational nucleus: non-degenerate low-lying $0^+, 2^+, 4^+$ states. We also
obtain the ground state binding energy well within the 1-percent level (1.6 MeV
/ 676 MeV). Notice that in addition to the yrast band we also reproduce the
yrare $K=2$ band. The last panel, $m=1.0 d_n$, is equivalent to the full
configuration interaction (FCI) calculation. Unlike the next three benchmark
cases, it is practical to compute the FCI results with PANASh, since the
dimensions are relatively modest ($10^5$).

\begin{figure*}[htb]
    \centering
    \includegraphics[width=.65\textwidth]{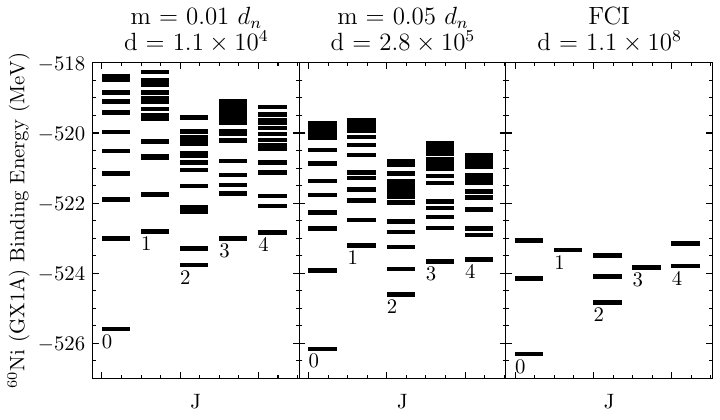}
    \caption[$^{60}$Ni Spectra]{Same as figure \ref{fig: ge} but for $^{60}$Ni.
    The final panel, labeled FCI, was performed with the $M$-scheme code {\tt
    BIGSTICK} at a dimension of $1.1\times 10^9$, but lists the equivalent
    $J$-scheme dimension. See text for a full discussion.}
    \label{fig: ni}
\end{figure*}

The results for $^{60}$Ni, an even-even spherical nucleus, are shown in
Fig.~\ref{fig: ni}. The format of the figure is the same as in Fig.~\ref{fig:
ge}, except for the last panel showing the FCI calculation. Here it was not
practical to get the FCI results using PANASh due to the large dimensions, and
instead the FCI code {\tt BIGSTICK} was used. Therefore, the dimension indicated
is in the $M$-scheme rather than the $J$-scheme (which would have been about an
order of magnitude smaller).
The spectral structure here is not so different from $^{78}$Ge, but we do see
seniority-like spectra as one might expect for a spherical, semi-magic
nucleus~\cite{ring2004nuclear}.
The ground state binding energies obtained using only $1\%$ of the basis factors
(4 orders of magnitude reduction in dimension) is within 1 MeV, and the
excitation energy of the first $2^+$ state is about 350 keV too high, about
twice the typical shell model uncertainty. Using $5\%$ of the basis factors
(still 3 orders of magnitude reduction), the ground state binding energy is
within 160 keV of FCI, and the first $2^+$ is within 81 keV.

\begin{figure*}[htb]
    \centering
    \includegraphics[width=.65\textwidth]{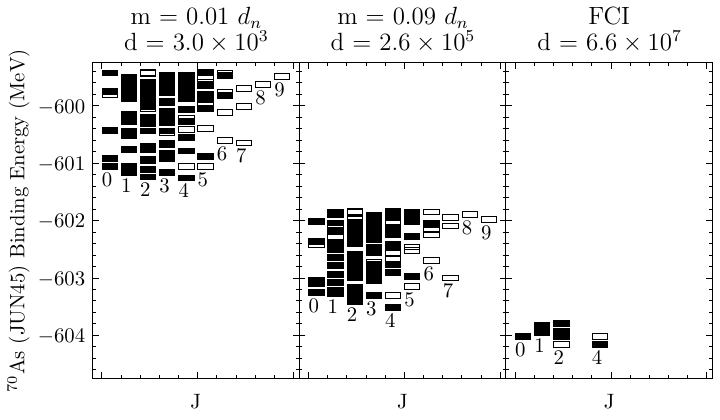}
    \caption[$^{70}$As spectra]{Same as figure \ref{fig: ge} but for $^{70}$As.
    The final panel, labeled FCI, was performed with the $M$-scheme code {\tt
    BIGSTICK} at a dimension of $7.1\times 10^8$, but lists the equivalent
    $J$-scheme dimension.}
    \label{fig: as}
\end{figure*}

The results for $^{70}$As, an odd-odd, deformed nucleus, are shown in
Fig.~\ref{fig: as}. 
Compared to  even-even nuclei with a $0^+$ ground state and orderly low-lying
excitations, the spectra for the odd-odd $^{70}$As is much denser with a $4^+$
ground state. While the $1\%$ calculation which in the previous two cases came
very close to the final ground state binding energy, here the discrepancy is
almost 3 MeV. This comports with our expectation that the odd proton and neutron
are forced to couple, increasing the proton neutron entanglement entropy and
reducing the effectiveness of the PANASh method. Despite this, we are still able
to obtain the approximate ordering of the low-lying states and obtain hundreds
of states where FCI can only manage a few with significant resources. For the
$9\%$ calculation, the $2^-$ ground state is off by 0.75 MeV and resolves above
the first $4^+$ MeV state which is nearly degenerate.

\begin{figure*}[htb]
    \centering
    \includegraphics[width=.65\textwidth]{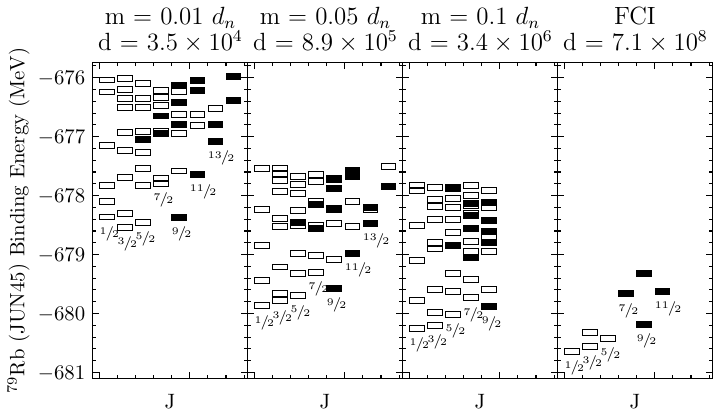}
    \caption[$^{79}$Rb spectra]{Same as Fig.~\ref{fig: ge} but for $^{79}$Rb. 
    The right-most panel was computed with the FCI code {\tt BIGSTICK} at a 
    dimension of $8.6 \times 10^9$, but lists the equivalent $J$-scheme dimension.}
    \label{fig: rb}
\end{figure*}

The results for $^{79}$Rb, an odd-A, spherical nucleus, are shown in
Fig.~\ref{fig: rb}. As an odd-A nucleus, there will be one unpaired nucleon
leading to half-integer spins (represented in the figure in decimal values).
The $M$-scheme dimension for this nucleus is $8.6\times 10^9$, which is
approaching the limits of our computing capabilities. Using {\tt BIGSTICK}, we
could only obtain the lowest four states of each parity in a reasonable amount
of time. In the $J$-scheme it would have been $7.1\times 10^8$. Using $1\%$ of
the basis factors, four orders of magnitude basis reduction, the ground state
binding energy is too high by 2.1 MeV, and the ordering of the first few states
does not match the converged results. This is not too surprising given the high
level density. Furthermore, this truncation error is comparable to the error of
FCI compared to the experimental binding energy: $\text{BE}_{exp}=-679.5$ MeV,
$\text{BE}_{FCI}=-680.7$ MeV (error $=1.2$ MeV). Increasing to $5\%$ of the basis
factors and an order of magnitude increase in dimension, we get the right
ordering of at least the first 5 states compared to FCI. The error in the ground
state binding energy is 770 keV. Doubling the basis factors to $10\%$ does not
significantly improve convergence despite quadrupling the basis dimension. This
is evidence that we can extract most of the physics from the lowest energy basis
factors.

The significant basis reduction achieved by the weak entanglement approximation
is useful in two extremes. The first, is that it makes possible calculations
which cannot be attempted in FCI - we will be able to study the structure of a
few low-lying states in model spaces that were previously computationally
impossible. The second extreme can already be seen by the small number of levels
in the FCI panel of last three figures: using this basis reduction method we can
obtain a far greater number of states for a comparable computational cost. We
can choose to sacrifice some quality for a larger quantity of states. This might
not sound desirable, but for statistical quantities such as average electromagnetic
properties of highly excited states, it is exactly the right trade-off.

\begin{figure}[ht!]
    \centering
    \includegraphics[width=0.5\textwidth]{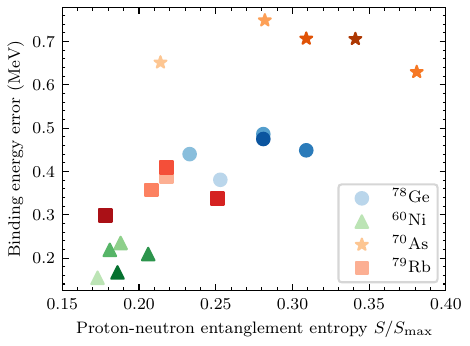}
    \caption{Error in the binding energy using a similar truncation for the five
    lowest states of the benchmark cases, as a function of the relative
    proton-neutron entanglement entropy. Nuclei with weaker entanglement tend to
    have a smaller error in the PANASh basis, as predicted
    in~\cite{johnson2023protonneutron}. Markers are shaded darker with
    increasing excitation energy; lower excited states tend to have lower
    entanglement. The binding energy and entanglements are computed from the
    largest truncation performed in FIG. 1 - 5 ($^{78}$Ge: $m=0.10 d_n$,
    $^{60}$Ni: $m=0.05 d_n$, $^{70}$As: $m=0.09 d_n$, $^{79}$Rb: $m=0.10 d_n$).
    The error is relative to the FCI (untruncated) values. The proton-neutron
    entanglement entropies are computed using the method described in
    section~\ref{sec:entangle} and as in~\cite{johnson2023protonneutron} and are
    divided by their maximum value in each model space to give a value between 0
    and 1.} \label{fig:err-vs-ent}
\end{figure}

Finally, we explored how the weak entanglement approximation is reflected in the
convergence properties of our four benchmark cases. Using the formalism
described in section~\ref{sec:entangle}, we computed the proton-neutron
entanglement entropy of the first five levels of each benchmark case. In
Fig.~\ref{fig:err-vs-ent} we show there is a correlation between the error in
the binding energy as computed with our PANASh code and the relative strength of
the proton-neutron entanglement (which we normalized to one). This matches our
expectations and the assumptions of the weak entanglement approximation. It
should be noted that the proton-neutron entanglement entropies were computed
using the approximate PANASh wave functions indicated in the caption of
Fig.~\ref{fig:err-vs-ent}, and are thus approximations themselves. 

\subsection{Aspirational calculation}\label{sec:heroic}

In the benchmark results presented above, at least a few of the low-lying states
can be extracted from the full shell model using state-of-the-art codes 
like~\cite{johnson2018bigstick}. In order to demonstrate the potential reach of
PANASh, we chose a case--$^{132}$Ce in the valence space bounded by magic
numbers 50 and 82, with a $^{100}$Sn core (that is, valence orbitals $0g_{7/2}$,
$2s_{1/2}$, $1d_{3/2}$, $1d_{5/2}$, $0h_{11/2}$)--whose FCI M-scheme dimension,
2.4 trillion, is far beyond current capabilities.  Because no FCI result is
possible, we compare to the experimental excitation
spectrum~\cite{KHAZOV2005497}. The interaction we use is tuned to tellurium
isotopes~\cite{PhysRevC.71.044317} and provides a reasonable description of
xenon isotopes~\cite{heimsoth2023uncertainties}. 
(Following~\cite{PhysRevC.71.044317}, we reduce the strength of the neutron-neutron
two-body matrix elements by 0.9.)  As a partial benchmark, we also compute in
this space with this interaction $^{128}$Xe, which has in this space an FCI
M-scheme dimension of 9.3 billion,  still tractable with suitable supercomputing
power; we take its experimental excitation energies from~\cite{ELEKES2015191}.

\begin{figure*}[htb!]
    \centering
    \includegraphics[width=0.7\textwidth, trim=0 3cm 0 3cm,clip]{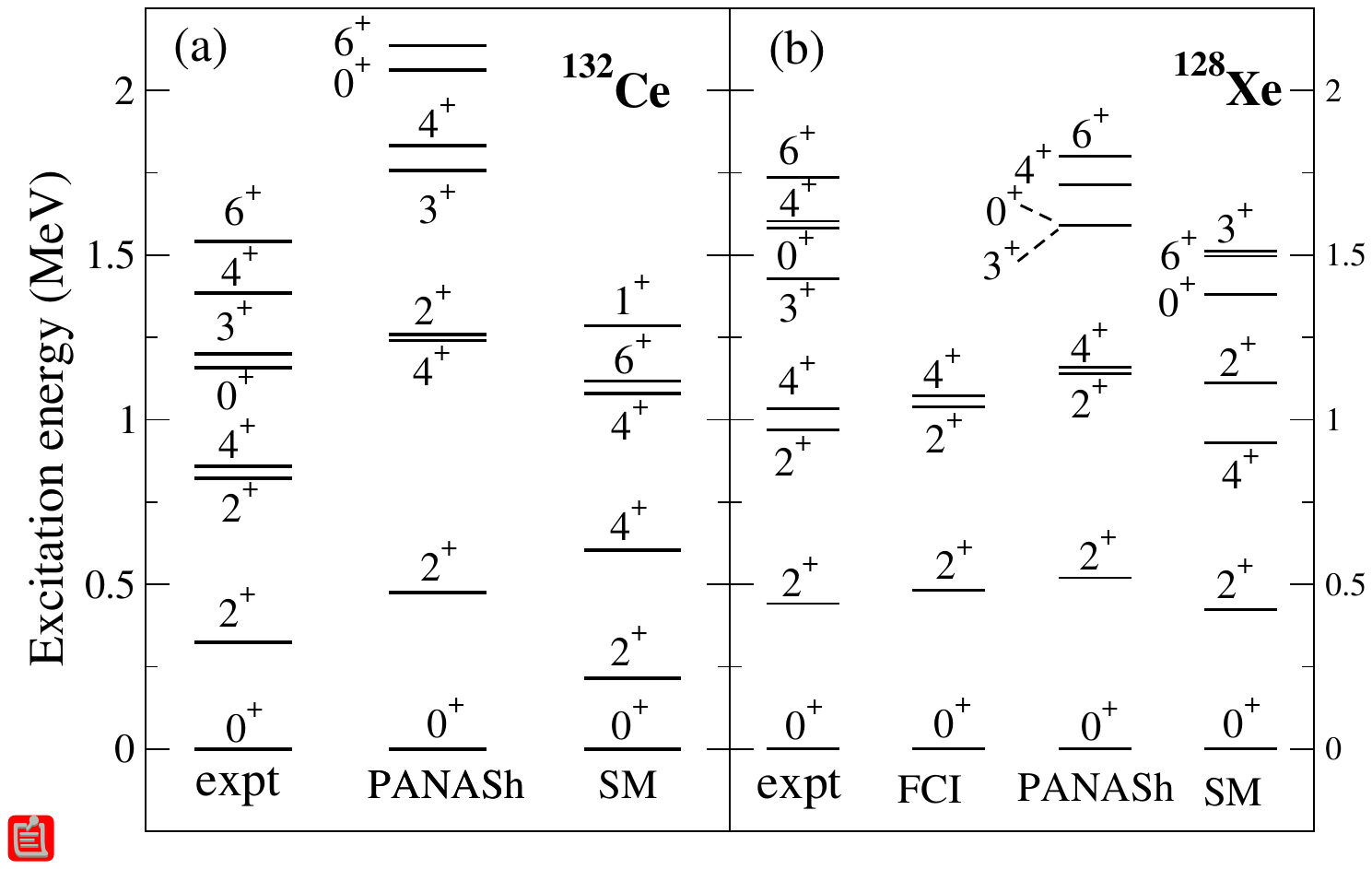}
    \caption{Excitation spectra for (a) $^{132}$Ce and (b) $^{128}$Xe, including
    from experiment (\cite{KHAZOV2005497} and~\cite{ELEKES2015191},
    respectively). We assume a $^{100}$Sn core and a valence space of the
    orbitals between magic numbers 50 and 82, and use the interaction 
    of~\cite{PhysRevC.71.044317}, including the recommended 0.9 reduction of
    neutron-neutron two-body matrix elements. Only for $^{128}$Xe is a full
    configuration-interaction (FCI) calculation possible, with an $M$-scheme
    dimension of 9.3 billion; the FCI $M$-scheme dimension for $^{132}$Ce would
    be 2.4 trillion. The PANASh calculations have neutron fractions $m=0.015 \,
    d_n$; the $J=6$ PANASh dimensions are 377\,000 for $^{128}$Xe and 432\,000 for
    $^{132}$Ce. Also shown are truncated (see text for details)
    configuration-interaction shell-model (SM) levels, with $M$-scheme
    dimensions of 507 million for $^{128}$Xe and 1.29 billion for $^{132}$Ce.}
    \label{fig:ce132xe128}
\end{figure*}

Fig.~\ref{fig:ce132xe128} presents the excitation spectra for $^{132}$Ce and
$^{128}$Xe.  In both cases we used 1\,000 proton and 1\,000 neutron levels to
construct the PANASh basis; this corresponds to a neutron basis fraction of  $m
= 0.015 \, d_n$, and proton basis fractions also $0.15 \, d_p$ for $^{132}$Ce
and $m =0.34 \, d_p$ for $^{128}$Xe. For $^{132}$Ce the largest PANSASh
$J$-scheme dimension used was 432\,000 for $J=6$, compared to the FCI $J=6$
dimension of 133 billion. For $^{128}$Xe, the corresponding PANASh $J=6$
dimension was 377\,000, compared to the FCI $J=6$ of 614 million. 

In addition to a calculation with PANASh, we carried out truncated calculation
with the {\tt BIGSTICK} code. We assign to each orbital $a$ an integer weight
$w_a$. Each many-body configuration is then assigned a total weight $W$ which is
the sum of the weights of the occupied orbitals.  The $M$-scheme Hilbert space
is truncated by keeping all configurations with weights up to a defined maximum
$W_\mathrm{max}$, defined relative to the minimum in the space $W_\mathrm{min}$:
$W_\mathrm{ex} = W_\mathrm{max} - W_\mathrm{min}$. With suitable choice of
weights, this truncation scheme~\cite{johnson2018bigstick} is flexible enough to
include the standard $N_\mathrm{max}$ truncation scheme for the no-core shell
model, as well as particle-hole truncations. We assigned the following weights
$w$ to each of the orbitals: $w(0g_{7/2}) = 0$; $w(1d_{5/2}) = 1$; $w(1d_{3/2})
= w(2s_{1/2}) = w(9h_{11/2}) = 2$. This choice of weights approximates a
truncation based upon the centroids (average energies) of orbital
configurations~\cite{PhysRevC.50.R2274,PhysRevC.90.024306}. Under this
truncation scheme with $W_\mathrm{ex}=6$,  $^{128}$Xe has an $M$-scheme
dimension of 507 million, while  $^{132}$Ce  has an $M$-scheme dimension of 1.29
billion. 

For $^{128}$Xe we get relatively good agreement for all three calculations as
well as with experiment. The truncated shell model (SM) calculation has a
slightly compressed yrast $0$-$2$-$4$ band, while the PANASh band is slightly
expanded. This behavior is exaggerated for $^{132}$Ce.  One can speculate that
the PANASh calculation, because it only partially couples the proton and neutron
sectors, underestimates the proton-neutron quadrupole collectivity, while the SM
calculation, which has reduced proton and neutron spaces, underestimates the
pairing collectivity in the ground state. We leave investigating this
speculation to future work.

Nonetheless, by comparing ground state energies we can demonstrate that PANASh
builds in substantial correlations. Table~\ref{tab:heroic} shows the ground
state energies obtained from our calculations. The absolute values of these
energies are not meaningful on their own, as the interaction was not fitted to
absolute binding energies, but they serve as a proxy for overall convergence due
to the variational principle of basis truncation methods. For $^{128}$Xe, the
truncated SM ground state is 1.53 MeV above the FCI ground state, but the PANASh
ground state is only 0.43 MeV above. While the FCI ground state energy for
$^{132}$Ce is not available, the PANASh ground state energy is 3.37 MeV lower
than the truncated SM ground state.
\begin{table*}[ht!]
    \caption{Ground state (GS) energies from three shell model calculations of
    $^{132}$Ce and $^{128}$Xe shown in Fig. ~\ref{fig:ce132xe128}. The value
    does not have a direct interpretation on its own since the interaction was
    not fit to the total binding energy. However, the variational principle
    guarantees that all truncations are less bound than FCI, and so the GS
    energy serves as a proxy for convergence. These values show that the PANASh
    calculation is more converged than the SM truncation while using a smaller
    basis.}\label{tab:heroic} 
    \centering
    \begin{tabular}{l c | c | c || c| c | c}
        \hline
                           &    \multicolumn{3}{c}{$^{132}$Ce}   & \multicolumn{3}{c}{$^{128}$Xe} \\
       Calculation         &   FCI & PANASh & SM trunc. & FCI & PANASh & SM trunc. \\
       \hline
       $J$-scheme (J=6) Dimension & $1.33\times 10^{11}$ & $4.32\times 10^{5}$ & - & $6.14\times 10^{8}$ & $3.77\times 10^{5}$ & - \\
       $M$-scheme Dimension & $2.4\times 10^{12}$ & - & $1.29\times10^{9}$ & $9.3\times 10^{9}$ & - & $5.07\times 10^{8}$ \\
       GS Energy (MeV)      & - & -291.17 & -287.79 & -264.41 & -263.98 & -262.88 \\
       First $2^+$ (MeV)    & - & -290.69 & -287.58 & -263.93 & -263.46 & -262.46
    \end{tabular}
\end{table*}

The FCI dimension for $^{132}$Ce is two orders of magnitude larger than any
published shell model calculation~\cite{forssen2018large,mccoy2024intruder}.
Although the PANASh excitation spectrum is not perfect, it captures the main
features of the experimental spectrum, and clearly builds in correlations beyond
what can be captured by a traditional truncated shell model calculation. As
discussed briefly in Section \ref{sec:outlook}, this will provide motivation to
push PANASh further and to carefully investigate extrapolation to the full space
with appropriate uncertainty quantification.

\section{Comparison to other SVD truncations}\label{sec:compare}

Our weak entanglement factorization has precursors in the density matrix
renormalization group (DMRG)~\cite{white1992density,
white1993densitymatrix,tichai2024spectroscopy,papenbrock2005density} and
variational wavefunction factorization (VWF) or simply wavefunction
factorization (as called by its creators~\cite{papenbrock2004solution,
papenbrock2003factorization}).  Both of these methods are based on a bipartite
decomposition of the Hilbert space, followed by some sort of truncation. Both
also directly or indirectly consider a singular-value decomposition of the
bipartite representation of the wave functions. 

In a DMRG approach to the nuclear shell
model, one splits the single-particle space into two subgroups which are
iteratively improved.
At each step of the DMRG method, (1) the effective interaction in each subspace
is diagonalized exactly, (2) the approximate ground state's SVD is used to
inform which states to keep in a truncation to $m$ factors from that subspace,
and (3) the solution to the full space is taken as a product state of the two
subspace solutions. Various flavors of DMRG deal with how to define the
subgroups, and how to change the subgroups at each iteration.

Variational wavefunction factorization  is also an iterative method which works
with a bipartite representation. A complete description can be found in the
literature by Papenbrock et al.~\cite{papenbrock2003factorization,
papenbrock2004solution}; a short summary will be given here. VWF seeks the
optimal set of proton and neutron factors $\ket{\tilde p}$ and $\ket{\tilde n}$
which for $m << \min (d_p,d_n)$ yield a good approximation,
\begin{equation}
    \ket{\Psi} \approx \sum_j^m  s_j \ket{\tilde p_j} \ket{\tilde n_j},
\end{equation}
It is known that the optimal factors are given by the SVD, but VWF deals with
the scenario where it is too computationally expensive to solve even the ground
state of the system. One starts with an ansatz state comprised of random proton
and neutron many-body wavefunctions and writes down a variational condition.
Then, one solves the coupled set of nonlinear equations that follow. This is
computed as a generalized eigenvalue problem. After each iteration, the number
of basis factors $m$ is increased until satisfactory convergence is reached.

Like DMRG, VWF relies on the fact that the singular values of realistic shell
model ground states fall off rapidly so that an accurate approximation can be
achieved with only a small number of factors~\cite{papenbrock2005density}.
Empirically, the spectra tends to converge exponentially with the number of
states $m$ retained~\cite{papenbrock2004solution}. 

\begin{figure*}[htb!]
    \centering
    \includegraphics[width=0.8\textwidth]{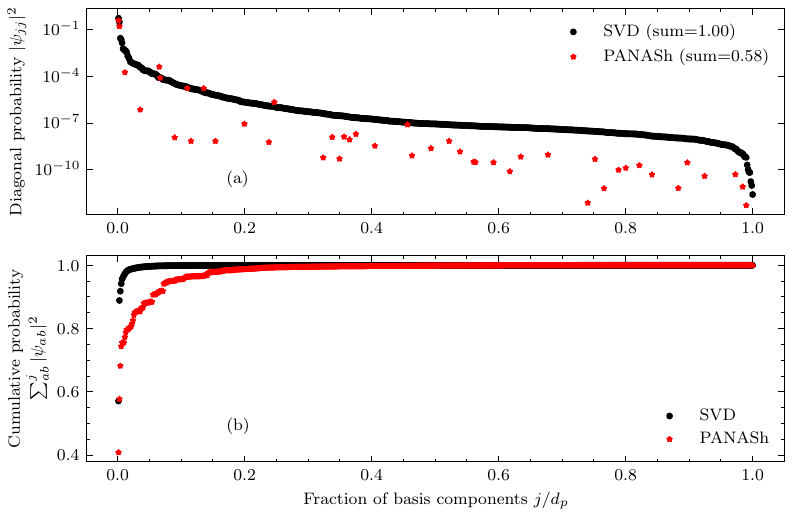}
    \caption{Representation of the (JUN45) $^{78}$Ge ground state wave function
    elements $\psi_{ij}$ in two different proton-neutron bases: the SVD basis in
    which the wave function is exactly diagonal, and the PANASh basis which for
    this nucleus is 58\% diagonal. Panel (a): diagonal elements (squared) of the
    wave function showing the PANASh basis follows the same exponentially
    decaying trend as the SVD basis. Panel (b): cumulative probability of the
    wave function captured as a function of the fraction of basis components.
    The PANASh basis is sub-optimal but the cumulative probability still
    converges to 1 exponentially in the number of basis components. This
    demonstrates that a significant truncation can be performed while
    maintaining a high fidelity wave function. }
    \label{fig:svd_vs_panash}
\end{figure*}

A major cost of both DMRG and VWF is the iterative approach to finding the
optimal basis factors. The weak entanglement approximation avoids this cost by
assuming that the eigenstates of the proton and neutron subgroups are sufficient
approximations of the optimal proton and neutron factors. Interestingly, our
choice of metric for selecting the proton and neutron basis factors is
reminiscent of Wilson's numerical renormalization group
(NRG)~\cite{wilson1975renormalization, dukelsky2004density}, the precursor to
DMRG. The downfall of NRG (which DMRG overcame) is that its truncation is based
only on the energy of the subspace solutions and ignores the coupling of the
subgroup to the rest of the system. It fails when long-range correlations exist.
Our method does not have this problem, since within each bipartition we include
all orbitals, and it is known that the entanglement between the proton and
neutron subgroups is weak~\cite{johnson2023protonneutron}. 

We can compare how close the weak entanglement approximation comes to the
optimal basis given by SVD. The SVD basis is the one where $\Psi = USV^T$ yields
a diagonal matrix $S$ whose elements are the singular values $s_i$. The values
$s_i^2$ are plotted and labeled ``SVD'' in Fig.~\ref{fig:svd_vs_panash} (a).
The PANASh basis is one where $\Psi = P \tilde{S} Q^T$ yields a matrix
$\tilde{S}$ (the proton-neutron wave function in the PANASh basis) which is
approximately diagonal, and with strongly decaying diagonal elements
$\psi_{jj}$. (The matrices $P$ and $Q$ would be formed by the proton and neutron
basis factors.) The values $|\psi_{jj}|^2$ are plotted for an untruncated PANASh
calculation and labeled ``PANASh'' in Fig.~\ref{fig:svd_vs_panash} (a). The SVD
matrix $S$ is completely diagonal and its diagonal elements sum to 1. The PANASh
matrix $\tilde{S}$ has off diagonal elements and only sums to 0.58 in this case.
The off-diagonal (0.42) strength is due to the fact that the PANASh basis is not
optimal for nonzero proton-neutron entanglement; for this nucleus the
proton-neutron entanglement is roughly 20\% of its maximum value. In panel (b)
of Fig.~\ref{fig:svd_vs_panash}, we plot the cumulative fraction of the wave
function as one includes more components, both of the ideal SVD basis and of the
more practical PANASh basis. While the SVD basis more quickly captures the full
wave function, as it must, the PANASh basis nonetheless provides a reasonable
approximation with significantly less numerical burden.

\section{\label{sec:outlook}Summary and Outlook}

We have presented a ``weak entanglement'' approximation to the nuclear shell
model, which relies upon the
observation~\cite{johnson2023protonneutron,perez-obiol2023quantum} that the
proton and neutron components, especially for neutron-rich nuclei, have
relatively low entanglement. Thus, rather than requiring computationally-expensive
optimization of the basis, as is done in precursor methods DMRG and VWF, we
simply construct the many-proton and many-neutron components independently as
eigenpairs of their respective Hamiltonians, an approach that arises from an
assumption of zero entanglement. If the entanglement were truly zero, the full
solutions would be simple tensor products; instead, we couple the proton and
neutron components through the proton-neutron interaction.  We find this
construction provides a `good enough' representation of the excitation spectra
of complex nuclei, including cases beyond current shell-model capabilities,
while retaining many of the advantages of configuration interaction, such as
generating many excited states, addressing open-shell and odd-particle systems,
and so on. 

There are a number of issues to be tackled in the immediate future, among them:
electromagnetic and weak transitions; improved basis construction; and
extrapolation and uncertainty quantification.  Basis construction might be
improved by relaxing the `zero entanglement' assumption: we could add to (for
example) the proton Hamiltonian a deformed field representing the average
influence of the neutron component, and vice versa.   Such work is under way and
will be reported on in a future paper.

\section*{Acknowledgements}

This work was performed in part under the auspices of the U.S. Department of
Energy by Lawrence Livermore National Laboratory under Contract
DE-AC52-07NA27344. Accordingly, the United States Government retains and the
publisher, by accepting the article for publication, acknowledges that the
United States Government retains a non-exclusive, paid-up, irrevocable,
world-wide license to publish or reproduce the published form of this article or
allow others to do so, for United States Government purposes. We acknowledge
additional support from the LLNL Weapon Physics and Design (WPD) Academic
Collaboration Team (ACT) University Collaboration program. This material is also
based upon work supported by the U.S. Department of Energy, Office of Science,
Office of Nuclear Physics, under Award Number DE-FG02-03ER41272. Part of this
research was enabled by computational resources supported by a generous gift to
SDSU from John Oldham. We would also like to thank Steven White for interesting
feedback on an early version of this work.

\appendix

\section{Residual interaction matrix elements}\label{sec:appendix}

To diagonalize the nuclear Hamiltonian $\h H = \h P + \h N + \h H^{(pn)}$ in the
basis Eq.~\eqref{base}, we seek an explicit form in terms of proton and neutron
one-body density matrices since these are a byproduct of the diagonalization of
the subspace Hamiltonians as defined in Eq.~\eqref{peig}.
The $\h P$ and $\h N$ operators will naturally be diagonal in our basis. (This
may be relaxed in the future.) The diagonal terms are $E^{(p)}$ and $E^{(n)}$.
All that remains is to find the matrix elements of $\h H^{(pn)}$ in terms of the
original FCI single particle basis and matrix elements. 

Starting from an explicit proton-neutron formalism, the (pn) part of the
interaction can be written in the form:
\begin{equation}\label{xpnh}
    \h H^{(pn)} = \sum_{abcd}\sum_K V_{ab,cd;K}^{(pn)}\sum_M 
    \h A_{ab;KM}^{\dagger(pn)}\h A_{cd;KM}^{(pn)},
\end{equation}
where subscripts $a$ and $c$ will refer to proton single-particle orbits, and
$b$ and $d$ to neutron single-particle orbits. The two-body operators are
defined as:
\begin{equation}
    \hat{A}_{ab;JM}^{\dagger(xy)}\equiv [x_a^\dagger \otimes y_b ^\dagger]_{JM}
    = \sum_{m_am_b} (j_am_a, j_bm_b|JM)\hat{x}^\dagger_{j_am_a}
        \hat{y}^\dagger_{j_bm_b}.
\end{equation}
Using various commutation relations and vector-coupling
identities~\cite{edmonds1996angular}, the following equivalent expression can be
derived:
\begin{align}\label{eq:hpn}
    H^{(pn)} = \sum_{abcd}\sum_{K}W^{(pn)}_{ac,bd;K}
    \hat{\rho}^{(p)}_{ac;K} \cdot \hat{\rho}^{(n)}_{bd;K}, 
\end{align}
where the one-body proton (neutron) density operators are defined as
\begin{equation}
   \h \rho^{(p)}_{ac;K\mu} \equiv \sum_{m_a m_c} (j_a m_a, j_c -m_c | K\mu) 
    \pi^\dagger_{j_am_a} \tilde{\pi}_{j_c-m_c},
\end{equation}
which carry total angular momentum $K$ and $z$-component $\mu$, and the
transformed proton-neutron interaction $W^{(pn)}$ is
\begin{align}
    W^{(pn)}_{ac,bd;K} 
    \equiv \sum_{K'} (-1)^{K'+j_b+j_c}[K']^2\begin{Bmatrix} j_a & j_b & K' \\ 
    j_d & j_c & K \end{Bmatrix}V_{ab,cd;K'}^{(pn)},
\end{align}
with $[x]=\sqrt{2x+1}$. Finally, the scalar-product of one-body density
operators is defined as
\begin{equation}
    \hat{\rho}^{(p)}_{ac;K} \cdot \hat{\rho}^{(n)}_{bd;K} 
    = \sum_\mu (-1)^{-\mu} 
    \h \rho^{(p)}_{ac;K\mu}
    \h \rho^{(n)}_{bd;K\mu}.
\end{equation}

The $H^{(pn)}$ part of the interaction is a scalar-product of two independent
operators: $\hat{\rho}^{(p)}_{ac;K} \cdot \hat{\rho}^{(n)}_{bd;K}$. Standard
vector-coupling relations allow us to write matrix elements of this operator in
a $j$-$j$-coupled basis as products of matrix elements in the uncoupled
basis~\cite{edmonds1996angular}:
\onecolumngrid
\begin{align}
    \langle f | \hat{\rho}^{(p)}_{ac;K} \cdot \hat{\rho}^{(n)}_{bd;K} | i \rangle  = 
    (-1)^{j_{pi}+j_{nf}+J} 
     \begin{Bmatrix}
    J & j_{nf} & j_{pf} \\
    K & j_{pi} & j_{ni} 
    \end{Bmatrix}
    \langle {p_f} || \hat{\rho}^{(p)}_{ac;K} || p_i \rangle 
    \langle {n_f} || \hat{\rho}^{(n)}_{bd;K} || n_i \rangle .
\end{align}
Here $| i \rangle = | p_in_i;J\pi \rangle$ and $(J\pi)_i=(J\pi)_f=J\pi$. The
matrix elements of the proton-neutron interaction is thus expressed in terms of
the one-body transition density matrix elements of the proton and neutron factor
wave functions.

The matrix elements of the pn-part of the Hamiltonian can be written in the
simplified form:
\begin{align}
    \langle f | \h H^{(pn)}_J | i \rangle
    =(-1)^{j_{pi}+j_{nf}+J}  
     \sum_{K} 
     \begin{Bmatrix}
    J & j_{nf} & j_{pf} \\
    K & j_{pi} & j_{ni} 
    \end{Bmatrix} 
    \sum_{bd}
    P_{bd;K}^{p_fp_i} \rho^{n_fn_i}_{bd;K}
\end{align}
where the reduced one-body density matrix elements of the factor wave functions
$| x \rangle$ are defined as
\begin{align}
     \rho_{ab;K}^{x_fx_i} \equiv 
     \langle x_f || \hat{\rho}^{(x)}_{ab;K} || x_i \rangle/\sqrt{2K+1},
\end{align}
and where the partial sum $P_{bd;K}^{p_fp_i}$ is defined as:
\begin{equation}\label{pstore}
\begin{split}
    P_{bd;K}^{p_fp_i} = (2K+1)
     \sum_{ac} 
    \rho^{p_fp_i}_{ac;K} W^{(pn)}_{ac,bd;K}.
\end{split}
\end{equation}
\twocolumngrid

\bibliographystyle{apsrev4-2}
\bibliography{library,panash}

\begin{thebibliography}{40}%
\makeatletter
\providecommand \@ifxundefined [1]{%
 \@ifx{#1\undefined}
}%
\providecommand \@ifnum [1]{%
 \ifnum #1\expandafter \@firstoftwo
 \else \expandafter \@secondoftwo
 \fi
}%
\providecommand \@ifx [1]{%
 \ifx #1\expandafter \@firstoftwo
 \else \expandafter \@secondoftwo
 \fi
}%
\providecommand \natexlab [1]{#1}%
\providecommand \enquote  [1]{``#1''}%
\providecommand \bibnamefont  [1]{#1}%
\providecommand \bibfnamefont [1]{#1}%
\providecommand \citenamefont [1]{#1}%
\providecommand \href@noop [0]{\@secondoftwo}%
\providecommand \href [0]{\begingroup \@sanitize@url \@href}%
\providecommand \@href[1]{\@@startlink{#1}\@@href}%
\providecommand \@@href[1]{\endgroup#1\@@endlink}%
\providecommand \@sanitize@url [0]{\catcode `\\12\catcode `\$12\catcode
  `\&12\catcode `\#12\catcode `\^12\catcode `\_12\catcode `\%12\relax}%
\providecommand \@@startlink[1]{}%
\providecommand \@@endlink[0]{}%
\providecommand \url  [0]{\begingroup\@sanitize@url \@url }%
\providecommand \@url [1]{\endgroup\@href {#1}{\urlprefix }}%
\providecommand \urlprefix  [0]{URL }%
\providecommand \Eprint [0]{\href }%
\providecommand \doibase [0]{https://doi.org/}%
\providecommand \selectlanguage [0]{\@gobble}%
\providecommand \bibinfo  [0]{\@secondoftwo}%
\providecommand \bibfield  [0]{\@secondoftwo}%
\providecommand \translation [1]{[#1]}%
\providecommand \BibitemOpen [0]{}%
\providecommand \bibitemStop [0]{}%
\providecommand \bibitemNoStop [0]{.\EOS\space}%
\providecommand \EOS [0]{\spacefactor3000\relax}%
\providecommand \BibitemShut  [1]{\csname bibitem#1\endcsname}%
\let\auto@bib@innerbib\@empty
\bibitem [{\citenamefont {Haxel}\ \emph {et~al.}(1949)\citenamefont {Haxel},
  \citenamefont {Jensen},\ and\ \citenamefont {Suess}}]{PhysRev.75.1766.2}%
  \BibitemOpen
  \bibfield  {author} {\bibinfo {author} {\bibfnamefont {O.}~\bibnamefont
  {Haxel}}, \bibinfo {author} {\bibfnamefont {J.~H.~D.}\ \bibnamefont
  {Jensen}},\ and\ \bibinfo {author} {\bibfnamefont {H.~E.}\ \bibnamefont
  {Suess}},\ }\href {https://doi.org/10.1103/PhysRev.75.1766.2} {\bibfield
  {journal} {\bibinfo  {journal} {Phys. Rev.}\ }\textbf {\bibinfo {volume}
  {75}},\ \bibinfo {pages} {1766} (\bibinfo {year} {1949})}\BibitemShut
  {NoStop}%
\bibitem [{\citenamefont {Mayer}(1949)}]{PhysRev.75.1969}%
  \BibitemOpen
  \bibfield  {author} {\bibinfo {author} {\bibfnamefont {M.~G.}\ \bibnamefont
  {Mayer}},\ }\href {https://doi.org/10.1103/PhysRev.75.1969} {\bibfield
  {journal} {\bibinfo  {journal} {Phys. Rev.}\ }\textbf {\bibinfo {volume}
  {75}},\ \bibinfo {pages} {1969} (\bibinfo {year} {1949})}\BibitemShut
  {NoStop}%
\bibitem [{\citenamefont {Kurath}(1956)}]{PhysRev.101.216}%
  \BibitemOpen
  \bibfield  {author} {\bibinfo {author} {\bibfnamefont {D.}~\bibnamefont
  {Kurath}},\ }\href {https://doi.org/10.1103/PhysRev.101.216} {\bibfield
  {journal} {\bibinfo  {journal} {Phys. Rev.}\ }\textbf {\bibinfo {volume}
  {101}},\ \bibinfo {pages} {216} (\bibinfo {year} {1956})}\BibitemShut
  {NoStop}%
\bibitem [{\citenamefont {Halbert}\ and\ \citenamefont
  {French}(1957)}]{PhysRev.105.1563}%
  \BibitemOpen
  \bibfield  {author} {\bibinfo {author} {\bibfnamefont {E.~C.}\ \bibnamefont
  {Halbert}}\ and\ \bibinfo {author} {\bibfnamefont {J.~B.}\ \bibnamefont
  {French}},\ }\href {https://doi.org/10.1103/PhysRev.105.1563} {\bibfield
  {journal} {\bibinfo  {journal} {Phys. Rev.}\ }\textbf {\bibinfo {volume}
  {105}},\ \bibinfo {pages} {1563} (\bibinfo {year} {1957})}\BibitemShut
  {NoStop}%
\bibitem [{\citenamefont {Whitehead}\ \emph {et~al.}(1977)\citenamefont
  {Whitehead}, \citenamefont {Watt}, \citenamefont {Cole},\ and\ \citenamefont
  {Morrison}}]{whitehead1977computationala}%
  \BibitemOpen
  \bibfield  {author} {\bibinfo {author} {\bibfnamefont {R.~R.}\ \bibnamefont
  {Whitehead}}, \bibinfo {author} {\bibfnamefont {A.}~\bibnamefont {Watt}},
  \bibinfo {author} {\bibfnamefont {B.~J.}\ \bibnamefont {Cole}},\ and\
  \bibinfo {author} {\bibfnamefont {I.}~\bibnamefont {Morrison}},\ }\bibinfo
  {title} {Computational methods for shell-model calculations},\ in\ \href
  {https://doi.org/10.1007/978-1-4615-8234-2_2} {\emph {\bibinfo {booktitle}
  {Advances in Nuclear Physics}}}\ (\bibinfo  {publisher} {Springer US},\
  \bibinfo {year} {1977})\ p.\ \bibinfo {pages} {123–176}\BibitemShut
  {NoStop}%
\bibitem [{\citenamefont {Forss\'en}\ \emph {et~al.}(2018)\citenamefont
  {Forss\'en}, \citenamefont {Carlsson}, \citenamefont {Johansson},
  \citenamefont {S\"a\"af}, \citenamefont {Bansal}, \citenamefont {Hagen},\
  and\ \citenamefont {Papenbrock}}]{forssen2018large}%
  \BibitemOpen
  \bibfield  {author} {\bibinfo {author} {\bibfnamefont {C.}~\bibnamefont
  {Forss\'en}}, \bibinfo {author} {\bibfnamefont {B.~D.}\ \bibnamefont
  {Carlsson}}, \bibinfo {author} {\bibfnamefont {H.~T.}\ \bibnamefont
  {Johansson}}, \bibinfo {author} {\bibfnamefont {D.}~\bibnamefont {S\"a\"af}},
  \bibinfo {author} {\bibfnamefont {A.}~\bibnamefont {Bansal}}, \bibinfo
  {author} {\bibfnamefont {G.}~\bibnamefont {Hagen}},\ and\ \bibinfo {author}
  {\bibfnamefont {T.}~\bibnamefont {Papenbrock}},\ }\href
  {https://doi.org/10.1103/PhysRevC.97.034328} {\bibfield  {journal} {\bibinfo
  {journal} {Phys. Rev. C}\ }\textbf {\bibinfo {volume} {97}},\ \bibinfo
  {pages} {034328} (\bibinfo {year} {2018})}\BibitemShut {NoStop}%
\bibitem [{\citenamefont {McCoy}\ \emph {et~al.}(2024)\citenamefont {McCoy},
  \citenamefont {Caprio}, \citenamefont {Maris},\ and\ \citenamefont
  {Fasano}}]{mccoy2024intruder}%
  \BibitemOpen
  \bibfield  {author} {\bibinfo {author} {\bibfnamefont {A.~E.}\ \bibnamefont
  {McCoy}}, \bibinfo {author} {\bibfnamefont {M.~A.}\ \bibnamefont {Caprio}},
  \bibinfo {author} {\bibfnamefont {P.}~\bibnamefont {Maris}},\ and\ \bibinfo
  {author} {\bibfnamefont {P.~J.}\ \bibnamefont {Fasano}},\ }\href@noop {}
  {\bibfield  {journal} {\bibinfo  {journal} {arXiv preprint arXiv:2402.12606}\
  } (\bibinfo {year} {2024})}\BibitemShut {NoStop}%
\bibitem [{\citenamefont {Hagen}\ \emph {et~al.}(2010)\citenamefont {Hagen},
  \citenamefont {Papenbrock}, \citenamefont {Dean},\ and\ \citenamefont
  {Hjorth-Jensen}}]{PhysRevC.82.034330}%
  \BibitemOpen
  \bibfield  {author} {\bibinfo {author} {\bibfnamefont {G.}~\bibnamefont
  {Hagen}}, \bibinfo {author} {\bibfnamefont {T.}~\bibnamefont {Papenbrock}},
  \bibinfo {author} {\bibfnamefont {D.~J.}\ \bibnamefont {Dean}},\ and\
  \bibinfo {author} {\bibfnamefont {M.}~\bibnamefont {Hjorth-Jensen}},\ }\href
  {https://doi.org/10.1103/PhysRevC.82.034330} {\bibfield  {journal} {\bibinfo
  {journal} {Phys. Rev. C}\ }\textbf {\bibinfo {volume} {82}},\ \bibinfo
  {pages} {034330} (\bibinfo {year} {2010})}\BibitemShut {NoStop}%
\bibitem [{\citenamefont {Otsuka}\ \emph {et~al.}(2001)\citenamefont {Otsuka},
  \citenamefont {Honma}, \citenamefont {Mizusaki}, \citenamefont {Shimizu},\
  and\ \citenamefont {Utsuno}}]{otsuka2001monte}%
  \BibitemOpen
  \bibfield  {author} {\bibinfo {author} {\bibfnamefont {T.}~\bibnamefont
  {Otsuka}}, \bibinfo {author} {\bibfnamefont {M.}~\bibnamefont {Honma}},
  \bibinfo {author} {\bibfnamefont {T.}~\bibnamefont {Mizusaki}}, \bibinfo
  {author} {\bibfnamefont {N.}~\bibnamefont {Shimizu}},\ and\ \bibinfo {author}
  {\bibfnamefont {Y.}~\bibnamefont {Utsuno}},\ }\href@noop {} {\bibfield
  {journal} {\bibinfo  {journal} {Progress in Particle and Nuclear Physics}\
  }\textbf {\bibinfo {volume} {47}},\ \bibinfo {pages} {319} (\bibinfo {year}
  {2001})}\BibitemShut {NoStop}%
\bibitem [{\citenamefont {Launey}\ \emph {et~al.}(2014)\citenamefont {Launey},
  \citenamefont {Dreyfuss}, \citenamefont {Draayer}, \citenamefont {Dytrych},\
  and\ \citenamefont {Baker}}]{launey2014emergence}%
  \BibitemOpen
  \bibfield  {author} {\bibinfo {author} {\bibfnamefont {K.~D.}\ \bibnamefont
  {Launey}}, \bibinfo {author} {\bibfnamefont {A.~C.}\ \bibnamefont
  {Dreyfuss}}, \bibinfo {author} {\bibfnamefont {J.~P.}\ \bibnamefont
  {Draayer}}, \bibinfo {author} {\bibfnamefont {T.}~\bibnamefont {Dytrych}},\
  and\ \bibinfo {author} {\bibfnamefont {R.}~\bibnamefont {Baker}},\
  }\href@noop {} {\bibfield  {journal} {\bibinfo  {journal} {Journal of
  Physics: Conference Series}\ }\textbf {\bibinfo {volume} {569}},\ \bibinfo
  {pages} {012061} (\bibinfo {year} {2014})}\BibitemShut {NoStop}%
\bibitem [{\citenamefont {McCoy}\ \emph {et~al.}(2020)\citenamefont {McCoy},
  \citenamefont {Caprio}, \citenamefont {Dytrych},\ and\ \citenamefont
  {Fasano}}]{PhysRevLett.125.102505}%
  \BibitemOpen
  \bibfield  {author} {\bibinfo {author} {\bibfnamefont {A.~E.}\ \bibnamefont
  {McCoy}}, \bibinfo {author} {\bibfnamefont {M.~A.}\ \bibnamefont {Caprio}},
  \bibinfo {author} {\bibfnamefont {T.~c.~v.}\ \bibnamefont {Dytrych}},\ and\
  \bibinfo {author} {\bibfnamefont {P.~J.}\ \bibnamefont {Fasano}},\ }\href
  {https://doi.org/10.1103/PhysRevLett.125.102505} {\bibfield  {journal}
  {\bibinfo  {journal} {Phys. Rev. Lett.}\ }\textbf {\bibinfo {volume} {125}},\
  \bibinfo {pages} {102505} (\bibinfo {year} {2020})}\BibitemShut {NoStop}%
\bibitem [{\citenamefont {Stumpf}\ \emph {et~al.}(2016)\citenamefont {Stumpf},
  \citenamefont {Braun},\ and\ \citenamefont {Roth}}]{PhysRevC.93.021301}%
  \BibitemOpen
  \bibfield  {author} {\bibinfo {author} {\bibfnamefont {C.}~\bibnamefont
  {Stumpf}}, \bibinfo {author} {\bibfnamefont {J.}~\bibnamefont {Braun}},\ and\
  \bibinfo {author} {\bibfnamefont {R.}~\bibnamefont {Roth}},\ }\href
  {https://doi.org/10.1103/PhysRevC.93.021301} {\bibfield  {journal} {\bibinfo
  {journal} {Phys. Rev. C}\ }\textbf {\bibinfo {volume} {93}},\ \bibinfo
  {pages} {021301} (\bibinfo {year} {2016})}\BibitemShut {NoStop}%
\bibitem [{\citenamefont {Dao}\ and\ \citenamefont
  {Nowacki}(2022)}]{PhysRevC.105.054314}%
  \BibitemOpen
  \bibfield  {author} {\bibinfo {author} {\bibfnamefont {D.~D.}\ \bibnamefont
  {Dao}}\ and\ \bibinfo {author} {\bibfnamefont {F.}~\bibnamefont {Nowacki}},\
  }\href {https://doi.org/10.1103/PhysRevC.105.054314} {\bibfield  {journal}
  {\bibinfo  {journal} {Phys. Rev. C}\ }\textbf {\bibinfo {volume} {105}},\
  \bibinfo {pages} {054314} (\bibinfo {year} {2022})}\BibitemShut {NoStop}%
\bibitem [{\citenamefont {Johnson}\ \emph {et~al.}(2013)\citenamefont
  {Johnson}, \citenamefont {Ormand},\ and\ \citenamefont
  {Krastev}}]{johnson2013factorization}%
  \BibitemOpen
  \bibfield  {author} {\bibinfo {author} {\bibfnamefont {C.~W.}\ \bibnamefont
  {Johnson}}, \bibinfo {author} {\bibfnamefont {W.~E.}\ \bibnamefont
  {Ormand}},\ and\ \bibinfo {author} {\bibfnamefont {P.~G.}\ \bibnamefont
  {Krastev}},\ }\href {https://doi.org/10.1016/j.cpc.2013.07.022} {\bibfield
  {journal} {\bibinfo  {journal} {Computer Physics Communications}\ }\textbf
  {\bibinfo {volume} {184}},\ \bibinfo {pages} {2761} (\bibinfo {year}
  {2013})}\BibitemShut {NoStop}%
\bibitem [{\citenamefont {Shimizu}(2013)}]{shimizu2013nuclear}%
  \BibitemOpen
  \bibfield  {author} {\bibinfo {author} {\bibfnamefont {N.}~\bibnamefont
  {Shimizu}},\ }\href@noop {} {\bibfield  {journal} {\bibinfo  {journal} {arXiv
  preprint arXiv:1310.5431}\ } (\bibinfo {year} {2013})}\BibitemShut {NoStop}%
\bibitem [{\citenamefont {Shimizu}\ \emph {et~al.}(2019)\citenamefont
  {Shimizu}, \citenamefont {Mizusaki}, \citenamefont {Utsuno},\ and\
  \citenamefont {Tsunoda}}]{shimizu2019thick}%
  \BibitemOpen
  \bibfield  {author} {\bibinfo {author} {\bibfnamefont {N.}~\bibnamefont
  {Shimizu}}, \bibinfo {author} {\bibfnamefont {T.}~\bibnamefont {Mizusaki}},
  \bibinfo {author} {\bibfnamefont {Y.}~\bibnamefont {Utsuno}},\ and\ \bibinfo
  {author} {\bibfnamefont {Y.}~\bibnamefont {Tsunoda}},\ }\href@noop {}
  {\bibfield  {journal} {\bibinfo  {journal} {Computer Physics Communications}\
  }\textbf {\bibinfo {volume} {244}},\ \bibinfo {pages} {372} (\bibinfo {year}
  {2019})}\BibitemShut {NoStop}%
\bibitem [{\citenamefont {Johnson}\ and\ \citenamefont
  {Gorton}(2023)}]{johnson2023protonneutron}%
  \BibitemOpen
  \bibfield  {author} {\bibinfo {author} {\bibfnamefont {C.~W.}\ \bibnamefont
  {Johnson}}\ and\ \bibinfo {author} {\bibfnamefont {O.~C.}\ \bibnamefont
  {Gorton}},\ }\href {https://doi.org/10.1088/1361-6471/acbece} {\bibfield
  {journal} {\bibinfo  {journal} {Journal of Physics G: Nuclear and Particle
  Physics}\ }\textbf {\bibinfo {volume} {50}},\ \bibinfo {pages} {045110}
  (\bibinfo {year} {2023})}\BibitemShut {NoStop}%
\bibitem [{\citenamefont {Pérez-Obiol}\ \emph {et~al.}(2023)\citenamefont
  {Pérez-Obiol}, \citenamefont {Masot-Llima}, \citenamefont {Romero},
  \citenamefont {Menéndez}, \citenamefont {Rios}, \citenamefont
  {García-Sáez},\ and\ \citenamefont
  {Juliá-Díaz}}]{perez-obiol2023quantum}%
  \BibitemOpen
  \bibfield  {author} {\bibinfo {author} {\bibfnamefont {A.}~\bibnamefont
  {Pérez-Obiol}}, \bibinfo {author} {\bibfnamefont {S.}~\bibnamefont
  {Masot-Llima}}, \bibinfo {author} {\bibfnamefont {A.~M.}\ \bibnamefont
  {Romero}}, \bibinfo {author} {\bibfnamefont {J.}~\bibnamefont {Menéndez}},
  \bibinfo {author} {\bibfnamefont {A.}~\bibnamefont {Rios}}, \bibinfo {author}
  {\bibfnamefont {A.}~\bibnamefont {García-Sáez}},\ and\ \bibinfo {author}
  {\bibfnamefont {B.}~\bibnamefont {Juliá-Díaz}},\ }\bibfield  {journal}
  {\bibinfo  {journal} {The European Physical Journal A}\ }\textbf {\bibinfo
  {volume} {59}},\ \href {https://doi.org/10.1140/epja/s10050-023-01151-z}
  {10.1140/epja/s10050-023-01151-z} (\bibinfo {year} {2023})\BibitemShut
  {NoStop}%
\bibitem [{\citenamefont {White}(1992)}]{white1992density}%
  \BibitemOpen
  \bibfield  {author} {\bibinfo {author} {\bibfnamefont {S.~R.}\ \bibnamefont
  {White}},\ }\href {https://doi.org/10.1103/PhysRevLett.69.2863} {\bibfield
  {journal} {\bibinfo  {journal} {Physical Review Letters}\ }\textbf {\bibinfo
  {volume} {69}},\ \bibinfo {pages} {2863} (\bibinfo {year}
  {1992})}\BibitemShut {NoStop}%
\bibitem [{\citenamefont {White}(1993)}]{white1993densitymatrix}%
  \BibitemOpen
  \bibfield  {author} {\bibinfo {author} {\bibfnamefont {S.~R.}\ \bibnamefont
  {White}},\ }\href {https://doi.org/10.1103/PhysRevB.48.10345} {\bibfield
  {journal} {\bibinfo  {journal} {Physical Review B}\ }\textbf {\bibinfo
  {volume} {48}},\ \bibinfo {pages} {10345} (\bibinfo {year}
  {1993})}\BibitemShut {NoStop}%
\bibitem [{\citenamefont {Papenbrock}\ and\ \citenamefont
  {Dean}(2005)}]{papenbrock2005density}%
  \BibitemOpen
  \bibfield  {author} {\bibinfo {author} {\bibfnamefont {T.}~\bibnamefont
  {Papenbrock}}\ and\ \bibinfo {author} {\bibfnamefont {D.~J.}\ \bibnamefont
  {Dean}},\ }\href@noop {} {\bibfield  {journal} {\bibinfo  {journal} {Journal
  of Physics G: Nuclear and Particle Physics}\ }\textbf {\bibinfo {volume}
  {31}},\ \bibinfo {pages} {S1377} (\bibinfo {year} {2005})}\BibitemShut
  {NoStop}%
\bibitem [{\citenamefont {Papenbrock}\ \emph {et~al.}(2004)\citenamefont
  {Papenbrock}, \citenamefont {Juodagalvis},\ and\ \citenamefont
  {Dean}}]{papenbrock2004solution}%
  \BibitemOpen
  \bibfield  {author} {\bibinfo {author} {\bibfnamefont {T.}~\bibnamefont
  {Papenbrock}}, \bibinfo {author} {\bibfnamefont {A.}~\bibnamefont
  {Juodagalvis}},\ and\ \bibinfo {author} {\bibfnamefont {D.~J.}\ \bibnamefont
  {Dean}},\ }\href {https://doi.org/10.1103/PhysRevC.69.024312} {\bibfield
  {journal} {\bibinfo  {journal} {Physical Review C: Nuclear Physics}\ }\textbf
  {\bibinfo {volume} {69}},\ \bibinfo {pages} {024312} (\bibinfo {year}
  {2004})}\BibitemShut {NoStop}%
\bibitem [{\citenamefont {Papenbrock}\ and\ \citenamefont
  {Dean}(2003)}]{papenbrock2003factorization}%
  \BibitemOpen
  \bibfield  {author} {\bibinfo {author} {\bibfnamefont {T.}~\bibnamefont
  {Papenbrock}}\ and\ \bibinfo {author} {\bibfnamefont {D.~J.}\ \bibnamefont
  {Dean}},\ }\href {https://doi.org/10.1103/PhysRevC.67.051303} {\bibfield
  {journal} {\bibinfo  {journal} {Physical Review C: Nuclear Physics}\ }\textbf
  {\bibinfo {volume} {67}},\ \bibinfo {pages} {051303} (\bibinfo {year}
  {2003})}\BibitemShut {NoStop}%
\bibitem [{\citenamefont {Andreozzi}\ and\ \citenamefont
  {Porrino}(2001)}]{andreozzi2001redundancyfree}%
  \BibitemOpen
  \bibfield  {author} {\bibinfo {author} {\bibfnamefont {F.}~\bibnamefont
  {Andreozzi}}\ and\ \bibinfo {author} {\bibfnamefont {A.}~\bibnamefont
  {Porrino}},\ }\href {https://doi.org/10.1088/0954-3899/27/4/309} {\bibfield
  {journal} {\bibinfo  {journal} {Journal of Physics G: Nuclear and Particle
  Physics}\ }\textbf {\bibinfo {volume} {27}},\ \bibinfo {pages} {845}
  (\bibinfo {year} {2001})}\BibitemShut {NoStop}%
\bibitem [{\citenamefont {Brown}\ and\ \citenamefont
  {Rae}(2014)}]{brown2014shell}%
  \BibitemOpen
  \bibfield  {author} {\bibinfo {author} {\bibfnamefont {B.}~\bibnamefont
  {Brown}}\ and\ \bibinfo {author} {\bibfnamefont {W.}~\bibnamefont {Rae}},\
  }\href@noop {} {\bibfield  {journal} {\bibinfo  {journal} {Nuclear Data
  Sheets}\ }\textbf {\bibinfo {volume} {120}},\ \bibinfo {pages} {115}
  (\bibinfo {year} {2014})}\BibitemShut {NoStop}%
\bibitem [{\citenamefont {Cole}(1999)}]{cole1999predicted}%
  \BibitemOpen
  \bibfield  {author} {\bibinfo {author} {\bibfnamefont {B.~J.}\ \bibnamefont
  {Cole}},\ }\href {https://doi.org/10.1103/PhysRevC.59.726} {\bibfield
  {journal} {\bibinfo  {journal} {Physical Review C: Nuclear Physics}\ }\textbf
  {\bibinfo {volume} {59}},\ \bibinfo {pages} {726} (\bibinfo {year}
  {1999})}\BibitemShut {NoStop}%
\bibitem [{\citenamefont {Honma}\ \emph {et~al.}(2009)\citenamefont {Honma},
  \citenamefont {Otsuka}, \citenamefont {Mizusaki},\ and\ \citenamefont
  {{Hjorth-Jensen}}}]{honma2009new}%
  \BibitemOpen
  \bibfield  {author} {\bibinfo {author} {\bibfnamefont {M.}~\bibnamefont
  {Honma}}, \bibinfo {author} {\bibfnamefont {T.}~\bibnamefont {Otsuka}},
  \bibinfo {author} {\bibfnamefont {T.}~\bibnamefont {Mizusaki}},\ and\
  \bibinfo {author} {\bibfnamefont {M.}~\bibnamefont {{Hjorth-Jensen}}},\
  }\href {https://doi.org/10.1103/PhysRevC.80.064323} {\bibfield  {journal}
  {\bibinfo  {journal} {Physical Review C: Nuclear Physics}\ }\textbf {\bibinfo
  {volume} {80}},\ \bibinfo {pages} {064323} (\bibinfo {year}
  {2009})}\BibitemShut {NoStop}%
\bibitem [{\citenamefont {Honma}\ \emph {et~al.}(2004)\citenamefont {Honma},
  \citenamefont {Otsuka}, \citenamefont {Brown},\ and\ \citenamefont
  {Mizusaki}}]{honma2004new}%
  \BibitemOpen
  \bibfield  {author} {\bibinfo {author} {\bibfnamefont {M.}~\bibnamefont
  {Honma}}, \bibinfo {author} {\bibfnamefont {T.}~\bibnamefont {Otsuka}},
  \bibinfo {author} {\bibfnamefont {B.~A.}\ \bibnamefont {Brown}},\ and\
  \bibinfo {author} {\bibfnamefont {T.}~\bibnamefont {Mizusaki}},\ }\href
  {https://doi.org/10.1103/PhysRevC.69.034335} {\bibfield  {journal} {\bibinfo
  {journal} {Physical Review C: Nuclear Physics}\ }\textbf {\bibinfo {volume}
  {69}},\ \bibinfo {pages} {034335} (\bibinfo {year} {2004})}\BibitemShut
  {NoStop}%
\bibitem [{\citenamefont {Ring}\ and\ \citenamefont
  {Schuck}(2004)}]{ring2004nuclear}%
  \BibitemOpen
  \bibfield  {author} {\bibinfo {author} {\bibfnamefont {P.}~\bibnamefont
  {Ring}}\ and\ \bibinfo {author} {\bibfnamefont {P.}~\bibnamefont {Schuck}},\
  }\href@noop {} {\emph {\bibinfo {title} {The Nuclear Many-Body Problem}}}\
  (\bibinfo  {publisher} {{Springer Science \& Business Media, New York}},\
  \bibinfo {year} {2004})\BibitemShut {NoStop}%
\bibitem [{\citenamefont {Johnson}\ \emph {et~al.}(2018)\citenamefont
  {Johnson}, \citenamefont {Ormand}, \citenamefont {McElvain},\ and\
  \citenamefont {Shan}}]{johnson2018bigstick}%
  \BibitemOpen
  \bibfield  {author} {\bibinfo {author} {\bibfnamefont {C.~W.}\ \bibnamefont
  {Johnson}}, \bibinfo {author} {\bibfnamefont {W.~E.}\ \bibnamefont {Ormand}},
  \bibinfo {author} {\bibfnamefont {K.~S.}\ \bibnamefont {McElvain}},\ and\
  \bibinfo {author} {\bibfnamefont {H.}~\bibnamefont {Shan}},\ }\href
  {https://doi.org/10.48550/ARXIV.1801.08432} {\bibinfo {title} {Bigstick: A
  flexible configuration-interaction shell-model code}} (\bibinfo {year}
  {2018})\BibitemShut {NoStop}%
\bibitem [{\citenamefont {Khazov}\ \emph {et~al.}(2005)\citenamefont {Khazov},
  \citenamefont {Rodionov}, \citenamefont {Sakharov},\ and\ \citenamefont
  {Singh}}]{KHAZOV2005497}%
  \BibitemOpen
  \bibfield  {author} {\bibinfo {author} {\bibfnamefont {Y.}~\bibnamefont
  {Khazov}}, \bibinfo {author} {\bibfnamefont {A.}~\bibnamefont {Rodionov}},
  \bibinfo {author} {\bibfnamefont {S.}~\bibnamefont {Sakharov}},\ and\
  \bibinfo {author} {\bibfnamefont {B.}~\bibnamefont {Singh}},\ }\href
  {https://doi.org/https://doi.org/10.1016/j.nds.2005.03.001} {\bibfield
  {journal} {\bibinfo  {journal} {Nuclear Data Sheets}\ }\textbf {\bibinfo
  {volume} {104}},\ \bibinfo {pages} {497} (\bibinfo {year}
  {2005})}\BibitemShut {NoStop}%
\bibitem [{\citenamefont {Brown}\ \emph {et~al.}(2005)\citenamefont {Brown},
  \citenamefont {Stone}, \citenamefont {Stone}, \citenamefont {Towner},\ and\
  \citenamefont {Hjorth-Jensen}}]{PhysRevC.71.044317}%
  \BibitemOpen
  \bibfield  {author} {\bibinfo {author} {\bibfnamefont {B.~A.}\ \bibnamefont
  {Brown}}, \bibinfo {author} {\bibfnamefont {N.~J.}\ \bibnamefont {Stone}},
  \bibinfo {author} {\bibfnamefont {J.~R.}\ \bibnamefont {Stone}}, \bibinfo
  {author} {\bibfnamefont {I.~S.}\ \bibnamefont {Towner}},\ and\ \bibinfo
  {author} {\bibfnamefont {M.}~\bibnamefont {Hjorth-Jensen}},\ }\href
  {https://doi.org/10.1103/PhysRevC.71.044317} {\bibfield  {journal} {\bibinfo
  {journal} {Phys. Rev. C}\ }\textbf {\bibinfo {volume} {71}},\ \bibinfo
  {pages} {044317} (\bibinfo {year} {2005})}\BibitemShut {NoStop}%
\bibitem [{\citenamefont {Heimsoth}\ \emph {et~al.}(2023)\citenamefont
  {Heimsoth}, \citenamefont {Lem}, \citenamefont {Suliga}, \citenamefont
  {Johnson}, \citenamefont {Balantekin},\ and\ \citenamefont
  {Coppersmith}}]{heimsoth2023uncertainties}%
  \BibitemOpen
  \bibfield  {author} {\bibinfo {author} {\bibfnamefont {D.~J.}\ \bibnamefont
  {Heimsoth}}, \bibinfo {author} {\bibfnamefont {B.}~\bibnamefont {Lem}},
  \bibinfo {author} {\bibfnamefont {A.~M.}\ \bibnamefont {Suliga}}, \bibinfo
  {author} {\bibfnamefont {C.~W.}\ \bibnamefont {Johnson}}, \bibinfo {author}
  {\bibfnamefont {A.~B.}\ \bibnamefont {Balantekin}},\ and\ \bibinfo {author}
  {\bibfnamefont {S.~N.}\ \bibnamefont {Coppersmith}},\ }\href
  {https://doi.org/10.1103/PhysRevD.108.103031} {\bibfield  {journal} {\bibinfo
   {journal} {Phys. Rev. D}\ }\textbf {\bibinfo {volume} {108}},\ \bibinfo
  {pages} {103031} (\bibinfo {year} {2023})}\BibitemShut {NoStop}%
\bibitem [{\citenamefont {Elekes}\ and\ \citenamefont
  {Timar}(2015)}]{ELEKES2015191}%
  \BibitemOpen
  \bibfield  {author} {\bibinfo {author} {\bibfnamefont {Z.}~\bibnamefont
  {Elekes}}\ and\ \bibinfo {author} {\bibfnamefont {J.}~\bibnamefont {Timar}},\
  }\href {https://doi.org/https://doi.org/10.1016/j.nds.2015.09.002} {\bibfield
   {journal} {\bibinfo  {journal} {Nuclear Data Sheets}\ }\textbf {\bibinfo
  {volume} {129}},\ \bibinfo {pages} {191} (\bibinfo {year}
  {2015})}\BibitemShut {NoStop}%
\bibitem [{\citenamefont {Horoi}\ \emph {et~al.}(1994)\citenamefont {Horoi},
  \citenamefont {Brown},\ and\ \citenamefont {Zelevinsky}}]{PhysRevC.50.R2274}%
  \BibitemOpen
  \bibfield  {author} {\bibinfo {author} {\bibfnamefont {M.}~\bibnamefont
  {Horoi}}, \bibinfo {author} {\bibfnamefont {B.~A.}\ \bibnamefont {Brown}},\
  and\ \bibinfo {author} {\bibfnamefont {V.}~\bibnamefont {Zelevinsky}},\
  }\href {https://doi.org/10.1103/PhysRevC.50.R2274} {\bibfield  {journal}
  {\bibinfo  {journal} {Phys. Rev. C}\ }\textbf {\bibinfo {volume} {50}},\
  \bibinfo {pages} {R2274} (\bibinfo {year} {1994})}\BibitemShut {NoStop}%
\bibitem [{\citenamefont {Jiao}\ \emph {et~al.}(2014)\citenamefont {Jiao},
  \citenamefont {Sun}, \citenamefont {Xu}, \citenamefont {Xu},\ and\
  \citenamefont {Qi}}]{PhysRevC.90.024306}%
  \BibitemOpen
  \bibfield  {author} {\bibinfo {author} {\bibfnamefont {L.~F.}\ \bibnamefont
  {Jiao}}, \bibinfo {author} {\bibfnamefont {Z.~H.}\ \bibnamefont {Sun}},
  \bibinfo {author} {\bibfnamefont {Z.~X.}\ \bibnamefont {Xu}}, \bibinfo
  {author} {\bibfnamefont {F.~R.}\ \bibnamefont {Xu}},\ and\ \bibinfo {author}
  {\bibfnamefont {C.}~\bibnamefont {Qi}},\ }\href
  {https://doi.org/10.1103/PhysRevC.90.024306} {\bibfield  {journal} {\bibinfo
  {journal} {Phys. Rev. C}\ }\textbf {\bibinfo {volume} {90}},\ \bibinfo
  {pages} {024306} (\bibinfo {year} {2014})}\BibitemShut {NoStop}%
\bibitem [{\citenamefont {Tichai}\ \emph {et~al.}(2024)\citenamefont {Tichai},
  \citenamefont {Kap{\'a}s}, \citenamefont {Miyagi}, \citenamefont {Werner},
  \citenamefont {Legeza}, \citenamefont {Schwenk},\ and\ \citenamefont
  {Zarand}}]{tichai2024spectroscopy}%
  \BibitemOpen
  \bibfield  {author} {\bibinfo {author} {\bibfnamefont {A.}~\bibnamefont
  {Tichai}}, \bibinfo {author} {\bibfnamefont {K.}~\bibnamefont {Kap{\'a}s}},
  \bibinfo {author} {\bibfnamefont {T.}~\bibnamefont {Miyagi}}, \bibinfo
  {author} {\bibfnamefont {M.}~\bibnamefont {Werner}}, \bibinfo {author}
  {\bibfnamefont {{\"O}.}~\bibnamefont {Legeza}}, \bibinfo {author}
  {\bibfnamefont {A.}~\bibnamefont {Schwenk}},\ and\ \bibinfo {author}
  {\bibfnamefont {G.}~\bibnamefont {Zarand}},\ }\href@noop {} {\bibfield
  {journal} {\bibinfo  {journal} {arXiv preprint arXiv:2402.18723}\ } (\bibinfo
  {year} {2024})}\BibitemShut {NoStop}%
\bibitem [{\citenamefont {Wilson}(1975)}]{wilson1975renormalization}%
  \BibitemOpen
  \bibfield  {author} {\bibinfo {author} {\bibfnamefont {K.~G.}\ \bibnamefont
  {Wilson}},\ }\href {https://doi.org/10.1103/RevModPhys.47.773} {\bibfield
  {journal} {\bibinfo  {journal} {Reviews of Modern Physics}\ }\textbf
  {\bibinfo {volume} {47}},\ \bibinfo {pages} {773} (\bibinfo {year}
  {1975})}\BibitemShut {NoStop}%
\bibitem [{\citenamefont {Dukelsky}\ and\ \citenamefont
  {Pittel}(2004)}]{dukelsky2004density}%
  \BibitemOpen
  \bibfield  {author} {\bibinfo {author} {\bibfnamefont {J.}~\bibnamefont
  {Dukelsky}}\ and\ \bibinfo {author} {\bibfnamefont {S.}~\bibnamefont
  {Pittel}},\ }\href {https://doi.org/10.1088/0034-4885/67/4/R02} {\bibfield
  {journal} {\bibinfo  {journal} {Reports on Progress in Physics}\ }\textbf
  {\bibinfo {volume} {67}},\ \bibinfo {pages} {513} (\bibinfo {year}
  {2004})}\BibitemShut {NoStop}%
\bibitem [{\citenamefont {Edmonds}(1996)}]{edmonds1996angular}%
  \BibitemOpen
  \bibfield  {author} {\bibinfo {author} {\bibfnamefont {A.~R.}\ \bibnamefont
  {Edmonds}},\ }\href@noop {} {\emph {\bibinfo {title} {Angular momentum in
  quantum mechanics}}}\ (\bibinfo  {publisher} {Princeton University Press},\
  \bibinfo {year} {1996})\BibitemShut {NoStop}%
\end{thebibliography}%
\end{document}